\documentclass[a4paper]{article}
\usepackage{blindtext}
\usepackage{soul}
\usepackage{geometry}
\title{Comments on 
``Fast and scalable search of whole-slide images
via self-supervised deep learning''}
\author{Milad Sikaroudi$^1$,Mehdi Afshari$^1$,Abubakr Shafique$^{1,2}$,Shivam Kalra$^1$,H.R.Tizhoosh$^{1,2}$\\
\emph{$^1$ Kimia Lab, University of Waterloo, ON, Canada}\\
\emph{$^2$ Rhazes Lab, Mayo Clinic, Rochester, MN, USA}}
\begin{document}
\date{}
\maketitle
Chen et al. [Chen2022] recently published the article “\ul{Fast and scalable search of whole-slide images via
self-supervised deep learning}” in Nature Biomedical Engineering. The authors call their method “self-supervised
image search for histology”, short SISH. The paper is not easily readable, and many important details are buried under ambiguous descriptions.

\vspace{0.1in}
\textbf{Incremental modification of Yottixel --} Yottixel introduced the concept of ``mosaic'' through a customized clustering and selection process [Kalra2020a]. While Chen et al. frequently mention ``Yottixel'' and ``mosaic,'' they only acknowledge once that they have followed the Yottixel's mosaic generation process. This inadequate acknowledgment fails to give proper credit to Yottixel's contributions to their scheme. It is evident that SISH cannot function without the Yottixel mosaic. Unfortunately, Chen et al. do not sufficiently emphasize the reliance of SISH on the Yottixel mosaic.

The task of searching in archives of gigapixel WSIs, like any other big-data problem, requires a well-established computer science strategy: ``Divide and Conquer.'' The Yottixel mosaic serves as the essential ``Divide'' stage, breaking down the challenging problem of WSI processing into manageable parts, represented by a mosaic of patches. This concept has been extensively validated [Kalra2020b]. SISH has borrowed the crucial \emph{Divide} element from Yottixel, albeit citing the Yottixel paper, without adequately explaining the significance of the mosaic. 

\vspace{0.1in}
It is important to recognize that WSI search is only possible through the Divide approach, and without a new \emph{patching} algorithm, no new solution can be achieved.

\vspace{0.1in}
\textbf{Not referencing MinMax binarization --} Chen et al. state, ``The binarization process converts a continuous vector to a binary string by starting from $\infty$ and then traversing through all elements in the vector to compare the value of the current element with the next one. If the next value is smaller than the current, it assigns the current value to 0, and 1 otherwise.'' This description corresponds to the \emph{MinMax algorithm} [Tizhoosh2016], which was utilized in Yottixel [Kalra2020a] as a simple yet effective approach for computing a 1D approximation of feature gradients. However, Chen et al. neither reference the MinMax paper nor its initial use on deep features [Kumar2018]. Consequently, readers may be under the impression that this method is being introduced by Chen et al. It is worth noting that the MinMax method is a patented technology that has been commercially implemented [Tizhoosh2020]. 

\vspace{0.1in}
It is highly irresponsible and misleading to claim that your method is ``open source'' when crucial components of it (that are not your own) are patented and commercialized.

\vspace{0.1in}
\textbf{SISH is a misnomer --} Chen et al. employ the term ``self-supervised image search.'' However, if the usage of self-supervision in training a network does not introduce a new \emph{pretext task}, it may not qualify as a novel approach. In particular, SISH lacks a new loss function and seems to rely on conventional augmentation methods for training. Additionally, it does not propose any search-oriented loss function. It is important to clarify that the term ``self-supervised image search'' does not mean that the search algorithm supervises the search queries. Rather, it claims that the SISH paper employs embeddings acquired through self-supervised training. However, strictly speaking, autoencoders do not perform self-supervision. Consequently, the term ``self-supervised image search'' is misleading and a misnomer. It is highly unlikely that the authors were unaware of the appropriate AI terminology.  

\vspace{0.1in}
In light of these clarifications, it appears that the term `self-supervised' has been included in the title to enhance the perceived novelty for the general readership in the pathology community.

\vspace{0.35in}

\textbf{Embeddings --} Yottixel uses DenseNet features (pre-trained on natural images), it also mentions that other networks might be used. \ul{Combining DenseNet with an autoencoder} does not create a new search framework worthy of a new name. The main question remains why the original framework of Yottixel for using a pre-trained DenseNet has not been replaced with the trained autoenoder? Keeping DenseNet (and subsequent barcoding, and before that the Yottixel's mosaic) clearly shows the attempt to slight modify the original Yottixel chain without any major innovation. 

\vspace{0.1in}

\noindent \textbf{Questions and concerns about experiments/comparisons --}
\begin{itemize}
\item \textbf{Ranking search results} -- Modifying search results, such as through additional ranking and classification, constitutes a ``post-processing'' procedure regardless of the search engine employed. When comparing different search engines, it is imperative to employ the same pre- and post-processing methods for all search procedures. This holds particular significance since Chen et al. emphasize that ``\emph{the ranking algorithm plays a crucial role
in the success of SISH}''. The classification and sorting of search results is not a novel concept [Ebrahimian2020] and should have been applied to Yottixel's results as well. Consequently, the reported results, which claim outlandish improvements of 45\%, are neither fair nor likely to be reliable.

\item \textbf{Speed} -- using vEB trees [Boas1975] and other logarithmic data structures are quite trivial and part of any
efficient implementation [Friedman1977]. Talking about theoretical upper bounds, O(n) versus O(log n), 
seems to be in negligence of the reality of high-grade commercial implementation of any such engine. Claiming ``\emph{theoretical constant time}'', in light of massive memory usage by SISH is another attempt to increase the perceived novelty. Besides, any  
vector-based search can be implemented efficiently. For instance, in one implementation of
IrisCodes [Daugman2015] “800 trillion ($8\times 10^{14}$) cross-comparisons are performed every day” by smart
memory usage of “great speed of Exclusive-OR (XOR) IrisCode matching, which executes at millions/sec
per CPU single core”, at a time where GPUs and multi-core parallel processing were not available. 
\end{itemize}

\vspace{0.1in}
\noindent  Borrowing existing concepts and labeling the modified version with a new name is simply repackaging. Searching in medical archives is a formidable task that necessitates years of research. Ideas and innovations from all active researchers are crucial. Regrettably, \emph{Chen et al.'s work is essentially a repackaging of Yottixel under a different name}. While it is acceptable to draw inspiration from each other's work, it is crucial to acknowledge that repackaging unfairly redirects credits and recognition to the authors of the repackaged work. Upholding academic integrity and honesty entails recognizing that only the original authors of a work have the authority to rename its incremental modifications.

\section*{References}
\noindent [Chen2022] Chen, Chengkuan, Ming Y. Lu, Drew FK Williamson, Tiffany Y. Chen, Andrew J. Schaumberg, and Faisal
Mahmood. "Fast and scalable search of whole-slide images via self-supervised deep learning." Nature Biomedical
Engineering 6, no. 12 (2022): 1420-1434.

\noindent [Kalra2020a] Kalra, S. et al. (2020).
Yottixel–an image search engine for large archives of histopathology whole slide images. Medical Image Analysis, 65,
101757.

\noindent [Kalra2020b] Kalra, S. et al., 2020. Pan-cancer diagnostic consensus through searching archival histopathology
images using artificial intelligence. NPJ digital medicine, 3(1), p.31.

\noindent [Boas1975] Peter van Emde Boas: Preserving order in a forest in less than logarithmic time (Proceedings of the 16th
Annual Symposium on Foundations of Computer Science 10: 75-84, 1975)

\noindent [Friedman1977] Friedman, J. H., Bentley, J. L., \& Finkel, R. A. (1977). An algorithm for finding best matches in logarithmic
expected time. ACM Trans. on Mathematical Software, 3(3), 209-226.

\noindent [Ebrahimian2020] Ebrahimian, A. et al. (2020). Class-aware image
search for interpretable cancer identification. IEEE Access, 8, 197352-197362.

\noindent [Daugman2015] Daugman, J. Information theory and the iriscode. IEEE TIFS, 11(2), 400-409, 2015.

\noindent [Tizhoosh2016] Tizhoosh, Hamid R. et al. ``Minmax radon
barcodes for medical image retrieval.'' International Symposium on Visual Computing, pp. 617-627. Springer, Cham, 2016.

\noindent [Kumar2018] Kumar, M. D., Babaie, M., \& Tizhoosh, H. R. Deep barcodes for fast retrieval of histopathology
scans. Int. Joint Conf. on Neural Networks (IJCNN), pp. 1-8, 2018.

\noindent [Tizhoosh2020] Tizhoosh, H. R. (2020). U.S. Patent No. 10,628,736. Washington, DC: U.S. Patent and Trademark Office

\end{document}